\documentclass[preprint]{revtex4-1}

\usepackage{graphicx}
\usepackage{dcolumn} 
\usepackage{bm} 

\begin{document}

\title{ Non-minimal  $R^\beta F^2$-Coupled  Electromagnetic Fields to Gravity and Static, Spherically Symmetric Solutions  }

\author{Tekin Dereli}
\email{tdereli@ku.edu.tr}
 \affiliation{Department of Physics,
Ko\c{c}
University, 34450 Sar{\i}yer, \.{I}stanbul, Turkey}
\author{\"{O}zcan Sert}
 \email{sertoz@itu.edu.tr}
\affiliation{Department of Physics Engineering, \.{I}stanbul Technical University, 34469 Maslak, \.{I}stanbul, Turkey}

\date{\today}

\begin{abstract}

 \noindent 
 We   investigate     non-minimal $R^\beta F^2$-type couplings of electromagnetic fields to gravity.
 We derive 
the field equations  by a first order variational principle using the
method of Lagrange multipliers. Then we  present  various static,
spherically symmetric solutions describing the exterior fields in the vicinity of electrically charged massive bodies. 

  \begin{description}
 \item[PACS numbers]: 04.20.-q, 04.40.Nr, 04.50.Kd.
\end{description}
 \end{abstract}

\maketitle


\def\ba{\begin{eqnarray}}
\def\ea{\end{eqnarray}}
\def\w{\wedge}



\section{Introduction}

\noindent    
 Non-minimal  couplings between  electromagnetic fields and gravity  can be induced in regions of space-time such as  the neighborhood of black holes or neutron stars  where high intensities of electromagnetic  and/or gravitational fields are present.  In such cases,  gravitational fields behave as a non-linear medium in which the electromagnetic fields propagate.
On the other hand,  very intense  electromagnetic fields may cause new gravitational effects.

\noindent  Non-minimal couplings of the form   $RF^2$  
were investigated to obtain more information on  the relationship between space-time curvature and electric charge conservation  \cite{prasanna}, \cite{horndeski}.
They were also derived by 
    Kaluza-Klein reduction  of  five-dimensional
$R^2$-Lagrangians  to four dimensions \cite{buchdahl, muller-hoissen, dereli2}. 
Similar
non-minimal coupling terms  were   obtained 
by calculating  QED one-loop
polarization on a curved background \cite{drummond}.
   Recently, the behavior of the rotational velocities of test particles gravitating around galaxies was investigated in modified gravity  for  non-minimal matter couplings \cite{harko1,harko2, nojiri}.  Furthermore,
   the non-minimal couplings of gravity and electromagnetism were extended to   $I(R) F^2$  form   
to explain late-time cosmic acceleration \cite{bamba1, bamba2},    primordial magnetic field during the reheating epoch \cite{lambiase} and  rotation curves of galaxies\cite{dereli1}.


 Nevertheless, the non-minimal couplings  have not been investigated in sufficient detail as regards to applications such as the contributions to  the rotational curve of galaxies and Pioneer anomaly.
To gain more insights on such configurations in non-minimal models,  we proceed to investigate exact solutions of them for  spherically symmetric systems with   charge and mass. 
 In order to  find more general solutions which are compatible with observations from the solar system to cosmological scales,   we propose     $R^\beta F^2$-type couplings between gravity and electromagnetism through arbitrary real power of the curvature scalar.
 We  determine  the field
equations by a first order variational principle using the method
of Lagrange multipliers. Then  we present a wide range of exact static, spherically symmetric solutions
that depend on various values of $\beta$.
It is interesting to note that the solutions are asymptotically  flat  
 in general for values of  $\beta \neq 0$ within the range $ -\frac{1}{3} < \beta < 1$.

The non-minimal couplings in general give contributions both to the Maxwell and the Einstein field equations.
Contributions to the Maxwell equations can be associated with the magnetization and the polarization of a specific medium while  contributions to the Einstein equations 
may give important modifications to the space-time metric. Such effects, if there are any, may be used to explain some unexpected observations of gravity 
such as dark matter, dark energy and Pioneer anomaly considering only the electromagnetic and gravitational fields.

\section{Field Equations of the Extended   Einstein-Maxwell  Theory} \label{model}

\bigskip

\noindent We will obtain our field equations by a variational
principle from an action
\begin{equation}
        I[e^a,{\omega^a}_b,F] = \int_M{L} = \int_M{\mathcal{L}^*1}
        \nonumber
\end{equation}
where  $\{e^a\}$ and ${\{\omega^a}_b\}$ are the fundamental
gravitational field variables and   $F$ is the electromagnetic
field 2-form.  The space-time metric $g = \eta_{ab} e^a \otimes
e^b$ with signature $(-+++)$ and we fix the orientation by setting
$*1 = e^0 \w e^1 \w  e^2 \w e^3 $.  Torsion 2-forms $T^a$ and
curvature 2-forms $R^{a}_{\; \; b}$ of space-time are found from
the Cartan-Maurer structure equations
\begin{equation}
de^a + \omega^{a}_{\;\;b} \w e^b = T^a , \nonumber
\end{equation}
\begin{equation}
d\omega^{a}_{\;\;b} + \omega^{a}_{\;\;c} \w \omega^{c}_{\;\;b} =
R^{a}_{\;\;b} . \nonumber
\end{equation}
We consider the following  Lagrangian density 4-form which involves 
$R^{\beta} F^2$ terms to all orders in $\beta$:
 \ba\label{lag1}
  L =  \frac{1}{2\kappa^2} R*1  - \frac{1}{2} {\sum\limits_{n=0}^{\infty}}((a_0 R)^\beta )^n F\w *F
  \label{Lagrange}
  \ea
where $\kappa^2 $ is  Newton's universal gravitational coupling constant,  $a_0$ is a coupling constant with dimension $[L]^2$ and $\beta $ is a real number. 
 When we take $a_0=0$, we get back to the minimal Einstein-Maxwell theory.  If we take $a_0 \neq 0$ and assume that the condition $|(a_0R)^\beta| < 1$ is satisfied, we can write
 ${\sum\limits_{n=0}^{\infty}}((a_0R)^\beta)^n  = \frac{1}{1- (a_0 R)^\beta}  $.
The right hand side
diverges  as $\beta \rightarrow 0$ or $ R \rightarrow \frac{1}{a_0}$. Therefore, the values $\beta =0$ and $a_0R = 1$ should be avoided. 
Thus we replace the Lagrangian (\ref{lag1}) with 
 \ba\label{lag22}
  L =  \frac{1}{2\kappa^2} R*1 - \frac{1}{2(1-(a_0R)^\beta)}  F\w *F. 
   \label{Lagrange}
   \ea
which is meaningful even when the condition $|(a_0R)^\beta| < 1$ is not satisfied.
 
   The field equations are obtained by considering the independent variations of
   the action with respect to  $\{e^a\}$,
   ${\{\omega^a}_b\}$ and $\{F\}$.  The electromagnetic field components are read  from the expansion $F = \frac{1}{2} F_{ab} e^a \w e^b$.
We will choose  the unique metric-compatible, torsion-free
Levi-Civita connection. We impose this choice of the connection
through constrained variations  by the method of Lagrange
multipliers.
 We constrain   the electromagnetic
field 2-form F to be closed, that is; $dF = 0$,
and this is imposed by the variation of a Lagrange multiplier 2-form
$\mu$.
Therefore, we add to the above Lagrangian density  the
following constraint terms:
\begin{equation}
L_{C} = \left ( de^a + \omega^{a}_{\;\;b} \w e^b \right ) \w
\lambda_a + dF  \w \mu  \nonumber
\end{equation}
where   $\lambda_a$'s are  Lagrange multiplier 2-forms whose
variation imposes the zero-torsion constraint  $T^a=0$.

\noindent The infinitesimal variations of the total Lagrangian
density $L + L_C$ (modulo a closed form) is found to be
\begin{eqnarray}\label{generaleinsteinfe1}
&& \dot{L} +{\dot{L}_C} = \frac{1}{2 \kappa^2} \dot{e}^a \w R^{bc}
\w *e_{abc} +  \dot{e}^a \w  \frac{1}{2(1-(a_0R)^\beta)}   (\iota_a F \w *F - F \w \iota_a *F)   +  \dot{e}^a \w D \lambda_a \nonumber
\\ & &
 + \dot{e}^a \w  \frac{\beta a_0^\beta R^{\beta -1}}{(1-(a_0R)^\beta)^2}  (\iota_a R^b)(\iota_b F \w *F  + F \w \iota_b *F)   + \frac{1}{2} \dot{\omega}_{ab} \w  ( e^b
\w \lambda^a - e^a \w \lambda^b)
  \nonumber \\
& &
+ \dot{\omega}_{ab} \w  {\Sigma}^{ab} 
 -\dot{ F} \w   \frac{1}{1-(a_0R)^\beta} *F   +
\dot{\lambda}_a \w T^a   - \dot{F} \w d\mu .
\end{eqnarray}
where the angular momentum tensor   ${\Sigma}^{ab} $:
\begin{eqnarray}\label{sigmaab1}
 {\Sigma}^{ab} &=&  -\frac{ 1}{2} D  [ \frac{\beta a_0^\beta R^{\beta-1}}{(1-(a_0R)^\beta)^2} (  F^{ab} *F 
  +  F^b \wedge \imath^a * F-  F^a \wedge \imath^b* F 
-   F\wedge \imath^{ab}*F) ].
   \end{eqnarray}
We will use the abbreviations $ e^a \wedge e^b \wedge \cdots =
e^{ab\cdots}$, $\iota_aF =F_a, \  \  \iota_{ba} F =F_{ab}, $ \   $ \iota_a {R^a}_b =R_b, \  \   \iota_{ba} R^{ab}= R $.
 Lagrange multiplier 2-forms $\lambda_a$ are solved
from the connection variation equations  
 \begin{eqnarray}\label{lambdaaeb}
 e_a\w \lambda_b -  e_b \w \lambda_a = 2{\Sigma}_{ab}
 \end{eqnarray}
by applying the interior product operators  twice as
\begin{eqnarray}\label{lambdaaeb2}
\lambda^a &=&  2\imath_b   {\Sigma}^{ba}  +\frac{1}{2} 
\imath_{bc}  {\Sigma}^{cb}\wedge e^a.
\end{eqnarray}
When we substitute the $ \lambda_a$\rq{}s above into the co-frame equation, we find the  Einstein field equations for the non-minimal theory:
\begin{eqnarray}\label{einstein}
&&  \frac{1}{2 \kappa^2}  R^{bc}
\w *e_{abc} +   \frac{1}{2(1-(a_0R)^\beta)}  (\iota_a F \w *F - F \w \iota_a *F)   +  D ( 2\imath_b   {\Sigma}^{ba}  +\frac{1}{2} e^a \wedge
\imath_{bc}  {\Sigma}^{cb})  \nonumber
\\ & &
 +  \frac{\beta a_0^\beta R^{\beta-1}}{(1-(a_0R)^\beta)^2} (\iota_a R^b)(\iota_b F \w *F + F \w \iota_b *F) =0 .
\end{eqnarray}
The Maxwell equations read
\begin{equation}\label{maxwell1}
dF = 0 \quad , \quad d* \left(  \frac{1}{(1-(a_0R)^\beta)}  F \right ) = 0 .
\end{equation}

\noindent
 We may represent  the effects of non-minimal
couplings of the electromagnetic fields to gravity  through the
definition of a constitutive tensor. Thus, Maxwell's equations for an
electromagnetic field $F$ in an arbitrary medium can be written as
\ba dF = 0 \quad , \quad *d*G = J \ea where $G$ is called the
excitation 2-form and $J$ is the source electric current density
1-form. The effects of gravitation and electromagnetism on matter
are described by $G$ and $J$. 
We can  complete this system  using 
electromagnetic constitutive relations relating $G$ and $J$ to
$F$.  Here we consider only the source-free interactions, that is
$J=0$. Then, we can write a simple linear constitutive relation \ba G =
{\cal{Z}} (F) \ea where ${\cal{Z}}$ is a type-(2,2)-constitutive
tensor. For the above theory, we have 
\ba G =  \frac{ 1}{1-(a_0R)^\beta} F . \ea 
 We can identify the  polarization  1-form $\emph{p}  =   \frac{  (a_0 R)^{\beta}}{1-(a_0R)^\beta}\iota_{U} F$ and the magnetization 1-form
  $\emph{ m}   =  -  \frac{  (a_0 R)^{\beta}}{1-(a_0R)^\beta}\iota_{U}*F  $ related with the above theory where $U$ is a unit, time-like velocity vector field associated  with an inertial observer.
  One can find more  information about these concepts in \cite{dereli}.

 \section{Static, Spherically Symmetric Solutions }

\noindent We look for static, spherically symmetric  solutions to the field equations which are given by the metric
\begin{equation}\label{metric}
              g = -f(r)^2dt^2  +  f(r)^{-2}dr^2 + r^2d\theta^2 +r^2\sin^2(\theta) d \phi^2.
\end{equation}
We consider a static electric potential 1-form  $A =
h(r) dt $. Then,  electromagnetic field 2-form
 \begin{eqnarray}\label{electromagnetic}
 F  &=&  dA = h\rq{}dr \w dt   = Hdr \w dt 
\end{eqnarray}   
has a spherically symmetric electric field component only. 

\noindent We note that, for the minimal case  $a_0=0$, the Reissner-Nordstr\"{o}m metric
\begin{eqnarray} \label{rn}
g=-(1-\frac{2M}{r}+ \frac{\kappa^2q^2}{2r^2})dt^2+(1-\frac{2M}{r}+ \frac{\kappa^2q^2}{2r^2})^{-1}dr^2+ r^2d\theta^2 +r^2\sin^2(\theta) d \phi^2
\end{eqnarray}
and the Coulomb electric potential 
\begin{eqnarray}\label{maxwell11}
A=-\frac{q}{r}dt
\end{eqnarray}
is a solution of   (\ref{einstein}) and (\ref{maxwell1}).

\noindent 
After a lengthy calculation, the   non-minimally coupled ($a_0\neq 0$)
Einstein-Maxwell equations  (\ref{einstein}) and (\ref{maxwell1}) are   reduced   for the metric (\ref{metric}) and the electromagnetic 2-form (\ref{electromagnetic}).
We do not find it useful to write down the reduced field equations  in full detail, because they are  very long and complicated.
In order to avoid the problems of having higher order derivatives in our field equations, we restrict attention to those cases for which the Lagrange multipliers $\lambda_a=0$.
We calculate  
\begin{eqnarray}\label{lambda}
\lambda_0 =-\lambda_2=\lambda_3= f\frac{d}{dr}(H^2    \frac{ \beta a_0^\beta R^{\beta-1}}{(1-(a_0R)^\beta)^2}  ) ,\hskip 1 cm and \hskip 1 cm  \lambda^1=0
\end{eqnarray}
where the curvature scalar is given by
\begin{eqnarray}
-R= {{f^2}\rq{}}\rq{} +\frac{4{f^2}\rq{} }{r}  +\frac{2}{r^2}(f^2-1).
\end{eqnarray}
 Thus, we set 
 \begin{eqnarray}\label{lambda1}
 H^2    \frac{ \beta a_0^\beta R^{\beta-1}}{(1-(a_0R)^\beta)^2}   = constant.
\end{eqnarray}
The Maxwell equations (\ref{maxwell1}) can be immediately integrated to give 
   \begin{eqnarray}\label{maxwell2}
 \frac{1}{1-(a_0R)^\beta}  Hr^2 &=& q
\end{eqnarray}
  where  q is the electric charge determined by the Gauss  integral $$ \frac{1}{4\pi} {\int_{S^2}{* G }} =q .$$
  The expressions (\ref{lambda1}) and (\ref{maxwell2}) imply the following relation for consistency:
  \begin{eqnarray}\label{R}
   R=a_1 r^{\frac{4}{\beta-1}}\label{R}
\end{eqnarray}
   where $a_1$ is an integration constant to be  fixed. Then, it can be shown easily that the remaining Einstein field equations become
\begin{eqnarray}\label{dif2}
\frac{1}{2}\left( {{f^2}\rq{}}\rq{} -\frac{2}{r^2}(f^2-1) \right) \left( \frac{1}{\kappa^2}+ H^2  \frac{\beta a_0^\beta R^{\beta-1}}{(1-(a_0R)^\beta)^2}  \right) -\frac{qH}{r^2}&=&0 , \\
\frac{R}{2}(1 - \frac{\kappa^2  \beta a_0^\beta R^{\beta-1}}{(1-(a_0R)^\beta)^2} H^2 ) \label{dif31}
&=&0  .
\end{eqnarray}

The coupled equations (\ref{R})-(\ref{dif31})  are solved by 
\begin{eqnarray}\label{nonminimalsol}
f^2(r) &=& 1-\frac{C}{r}+\frac{ \kappa^2q^2} {4r^2}  - \frac{a_1(\beta-1)^2}{4\beta(3\beta+1)}r^{\frac{2\beta+2}{\beta-1}} 
 \hskip 1 cm for \ a_0 \neq 0, \  \  \beta \neq 0,1,-\frac{1}{3}  \nonumber \\
 h(r) &=& -\frac{q}{r} -\frac{a_1(\beta -1)}{ \kappa^2 q \beta (3\beta + 1)}r^{\frac{3\beta+1}{\beta-1}} .   
\end{eqnarray}
with  $a_1= (\kappa^2 q^2 \beta a_0^\beta )^{\frac{1}{1-\beta}}$. If we  go back to non-minimally coupled  Einstein-Maxwell field equations and check the solutions, 
we fix  the integration constant  in the (\ref{lambda1}) 
to be $\frac{1}{\kappa^2}$. We further note that the values of $\beta\neq 0$ in the range 
 $-\frac{1}{3} <\beta <1 $ give asymptotically flat solutions.  For $\beta$ in this interval, we identify the mass
 $M=\frac{C}{2}$.
 Moreover, $r=0$ is an essential singularity in general where  the quadratic curvature invariant $*(R_{ab} \wedge * R^{ab})\rightarrow \infty $  as $r \rightarrow 0$.
It is remarkable to observe that  the solution (\ref{nonminimalsol})  goes to the Schwarzschild solution for $q=0$, while for $q \neq 0$ but $a_0 = 0$ the metric in (\ref{nonminimalsol})  does not 
coincide with the Reissner-Nordstr\"{o}m metric (\ref{rn}).  The explanation is as follows: For $a_0 =0$, we have $R=0$, so that (\ref{dif31}) is identically satisfied. In this case, (\ref{dif2}) is solved by the Reissner-Nordstr\"{o}m metric. On the other hand, for $a_0 \neq 0$ and $a_1\neq 0$ we solve (\ref{dif31}) with the choice 
$$
\frac{\kappa^2  \beta a_0^\beta R^{\beta-1}}{(1-(a_0R)^\beta)^2} H^2 = 1
$$
for which the limit $a_0 \rightarrow 0$ does not exist.  
We would like to point out that with the choice
$$
\frac{\kappa^2  \beta a_0^\beta R^{\beta-1}}{(1-(a_0R)^\beta)^2} H^2 = 0
$$
for an arbitrary $a_0 \neq 0 $, we may set $a_1 = 0$ in (\ref{R}) and show that our non-minimally coupled model admits the Reissner-Nordstr\"{o}m black hole  (\ref{rn}) and (\ref{maxwell11}) as a solution.
 
   \bigskip
\noindent 
A  metric function similar to ours above with  $\beta =\frac{1}{3}$ is obtained  and analyzed in a recent paper \cite{balakin}.  They  consider a different theory of  non-minimally coupled electromagnetic fields  to gravity. Furthermore, their electromagnetic field is that of a Dirac  magnetic monopole $Q_m$.  Contrary to our  case, this solution
 joins smoothly to the Reissner-Nordstr\"{o}m solution in the absence of non-minimal couplings. 
We also wish to point out that the  metric (\ref{nonminimalsol})  with $\beta =- 3$ 
  involves a Rindler acceleration term.  This is   noted by  
\cite{grumiller, grumiller2}  in a dilaton-gravity model and 
may be  related with the  source of  various anomalies such as  the rotation curves of spiral galaxies and the Pioneer anomaly.

 
\section{ Conclusion}

\noindent We have formulated  non-minimal
$R^\beta F^2$-type coupled   Einstein-Maxwell theory. The field equations are obtained  by a first order variational
principle using the method of Lagrange multipliers in the language
of exterior differential forms.  We found  static,
spherically symmetric solutions for  arbitrary values of the $\beta$ parameter.  
Such solutions may be related with the source of various anomalies in  gravity  for certain values of the parameters $a_0$ and $\beta$. Moreover,  they   may contribute to resolution of    many important challenges  such as dark matter, dark energy and Pioneer anomaly without any other fields.
In addition, the non-minimal terms modify electromagnetic potentials at cosmological scales. Then the conventional electromagnetic energy density of the universe gets modified.
In other words; if the effects of dark matter are not due to some exotic matter fields, the non-minimal couplings can lead to such effects\cite{nojiri4}. 
Even if the electric charge q is not large,  one may choose the parameters in such a way that  these solutions can explain the rotation curves of galaxies and Pioneer anomaly for some parameter values.

\vskip 1cm

\section{Acknowledgement}

\noindent  T.D. gratefully acknowledges partial  support from The
Turkish Academy of Sciences (TUBA) and
 \"{O}.S. would like to  acknowledge fruitful discussions with Muzaffer Adak.

\vskip 1cm


\end{document}